\documentstyle[multicol,epsf,aps,pre]{revtex}

\begin{document}
\title{Fitness versus Longevity in Age-Structured Population Dynamics}
\author{W. Hwang, P.~L.~Krapivsky, and S. Redner}
\address{Center for BioDynamics, Center for Polymer Studies, 
and Department of Physics, Boston University, Boston, MA 02215}

\maketitle 
\begin{abstract} 
  We examine the dynamics of an age-structured population model in which the
  life expectancy of an offspring may be mutated with respect to that of the
  parent.  While the total population of the system always reaches a steady
  state, the fitness and age characteristics exhibit counter-intuitive
  behavior as a function of the mutational bias.  By analytical and numerical
  study of the underlying rate equations, we show that if deleterious
  mutations are favored, the average fitness of the population reaches a
  steady state, while the average population age is a {\em decreasing}
  function of the overall fitness.  When advantageous mutations are favored,
  the average population fitness grows linearly with time $t$, while the
  average age is independent of fitness.  For no mutational bias, the average
  fitness grows as $t^{2/3}$.

\end{abstract}

\begin{multicols}{2}
\narrowtext


\section{Introduction}

The goal of this paper is to understand the role of mutations on the
evolution of fitness and age characteristics of individuals in a simple
age-structured population dynamics model\cite{hkr}.  While there are many
classical models to describe single-species population
dynamics\cite{murray,nisbet}, consideration of age-dependent characteristics
is a more recent development\cite{murray,charlesworth,metz,penna,evolution}.
Typically, age characteristics of a population are determined by studying
rate equations which include age-dependent birth and death rates.  Here we
will study an extension of age-structured population dynamics in which the
characteristics of an offspring are mutated with respect to its parent.  In
particular, an offspring may be more ``fit'' or less fit than its parent, and
this may be reflected in attributes such as its birth rate and/or its life
expectancy.

\begin{figure}
\narrowtext
\epsfxsize=50mm\epsfysize=40mm
\hskip 0.25in\epsfbox{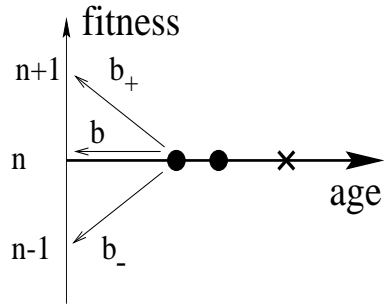}
\vskip 0.1in
\caption{Schematic illustration of the model.  An individual with fitness, or
  intrinsic life expectancy $n$, continuously ages (horizontal arrow).  The
  heavy dots signify individual birth events.  At each birth event, an
  offspring (of age zero) is produced, whose intrinsic lifetime is either
  $n+1$, $n$, or $n-1$, with relative rates $b_+$, $b$, and $b_-$,
  respectively.  The individual dies either by aging or by competition
  ($\times$).
\label{scheme}}
\end{figure}

In our model, we characterize the fitness of an individual by a single
heritable trait -- the life expectancy $n$ -- which is defined as the average
life span of an individual in the absence of competition.  This provides a
useful fitness measure, as a longer-lived individual has a higher chance of
producing more offspring throughout its life span.  We allow for either
deleterious or advantageous mutations, where the offspring fitness is less
than or greater than that of the parent, respectively (Fig.~\ref{scheme}).
This leads to three different behaviors which depend on the ratio between
these two mutation rates.  When advantageous mutation is favored, the fitness
distribution of the population approaches a Gaussian, with the average
fitness growing linearly with time $t$ and the width of the distribution
increasing as $t^{1/2}$.  Conversely, when deleterious mutation is more
likely, a steady-state fitness distribution is approached, with the rate of
approach varying as $t^{-2/3}$.  When there is no mutational bias, the
fitness distribution again approaches a Gaussian, but with average fitness
growing as $t^{2/3}$ and the width of the distribution again growing as
$t^{1/2}$.

In all three cases, the average population age reaches a steady state which,
surprisingly, is a {\em decreasing} function of the average fitness.  Thus
within our model, {\em a more fit population does not lead to an increased
  individual lifetime}.  Qualitatively, as individuals become more fit,
competition plays a more prominent role and is the primary mechanism that
leads to premature mortality.

In the following two sections, we formally define the model and outline
qualitative features of the population dynamics.  In Secs.~IV-VI, we analyze
the three cases of deleterious, advantageous, and neutral mutational biases
in detail.  We conclude in Sec.~VII.  Various calculational details are
provided in the Appendices.


\section{The Model}

Our model is a simple extension of logistic dynamics in which a population
with overall density $N(t)$ evolves both by birth at rate $b$ and death at
rate $\gamma N$.  Such a system is described by the rate equation
\begin{equation}
  \label{rate}
\dot N=b N-\gamma N^2,
\end{equation}
with steady-state solution $N_\infty=b/\gamma$.  Our age-structured mutation
model incorporates the following additional features:
\begin{enumerate}
  
\item Each individual is endowed with given life expectancy $n$.  This means
  that an individual has a rate of dying which equals $1/n$.
  
\item Death by aging occurs at a rate inversely proportional to the life
  expectancy.
  
\item Individuals give birth at a constant rate during their lifetimes.
  
\item In each birth event, the life expectancy of the newborn may be equal
  to that of its parent, or the life expectancy may be increased by or decreased
  by 1.  The relative rates of these events are $b$, $b_+$, and $b_-$,
  respectively.
\end{enumerate}

Each of these features represent idealizations.  Most prominently, it would
be desirable to incorporate a realistic mortality rate which is an increasing
function of age\cite{charlesworth,evolution,azbel}.  However, we argue in
Sec.~VII that our basic conclusions continue to be valid for systems with
realistic mortality rates.

To describe this dynamics mathematically, we study $C_n(a,t)$, the density of
individuals with life expectancy $n\ge 1$ and age $a$ at time $t$.  We also
introduce $P_n(t)=\int_0^\infty C_n(a,t)\, da$, which is the density of
individuals with life expectancy $n$ and {\em any} age at time $t$.  Finally,
the total population density is the integral of the population density over
all ages and life expectancies,
\begin{equation}
  \label{def}
N(t) = \sum_{n=1}^\infty \int_0^\infty  C_n(a,t)\, da= \sum_{n=1}^\infty P_n(t).
\end{equation}

According to our model, the rate equation for $C_n(a,t)$ is
\begin{equation}
\label{cn}
\left({\partial \over\partial t}+{\partial \over\partial a}\right)C_n(a,t)
=-\left(\gamma N(t)+{1\over n}\right)C_n(a,t). 
\end{equation}
The derivative with respect to $a$ on the left-hand side accounts for the
continuous aging of the population\cite{murray,charlesworth}.  On the 
right-hand side, $\gamma NC_n$ is the death rate due to competition, which is
assumed to be independent of an individual's age and fitness.  As discussed
above, the mortality rate is taken as age independent, and the form $C_n/n$
guarantees that the life expectancy in the absence of competition equals $n$.
Because birth creates individuals of age $a=0$, the population of newborns
provides the following boundary condition for $C_n(0,t)$,
\begin{equation}
\label{boundary}
C_n(0,t)=bP_n(t)+ b_+P_{n-1}(t)+ b_-P_{n+1}(t).
\end{equation}
Finally, the condition $P_0=0$ follows from the requirement that offspring
with zero life expectancy cannot be born.

\section{Basic Population Characteristics}

Let us first study the fitness characteristics of the population and
disregard the age structure.  The rate equation for $P_n(t)$ is found by
integrating Eq.~(\ref{cn}) over all ages and then using the boundary
condition Eq.~(\ref{boundary}) to give
\begin{equation}
\label{pn}
{d P_n\over dt}=\left(b-\gamma N-{1\over n}\right)P_n 
+ b_+P_{n-1}+ b_-P_{n+1}.
\end{equation}
This describes a random-walk-like process with state-dependent hopping rates
in the one-dimensional fitness space $n$.  Notice the hidden non-linearity
embodied by the term $\gamma NP_n$, since the total population density
$N(t)=\sum_n P_n(t)$.  From Eq.~(\ref{pn}), we find that $N(t)$ obeys a
generalized logistic equation
\begin{equation}
\label{N}
{d N\over dt}=(b+ b_++ b_--\gamma N)N-\sum_{n=1}^\infty{P_n\over n} 
- b_-P_1.
\end{equation}

\begin{figure}
\narrowtext
\epsfxsize=75mm\epsfysize=25mm
\hskip 0.15in\epsfbox{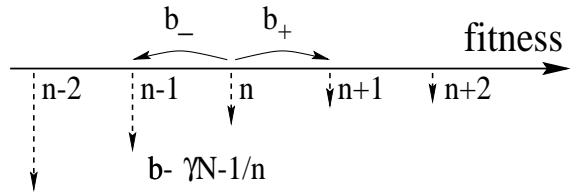}
\vskip 0.1in
\caption{Illustration of the random walk in fitness space that underlies
  the behavior of $P_n$.  The term $(b-\gamma N-{1\over n})P_n<0$ represents
  a state-dependent population loss (dashed arrows) which decreases for
  larger $n$.  
\label{Pn}}
\end{figure}

The three different dynamical regimes outlined in the introduction are
characterized by the relative magnitudes of the mutation rates $b_+$ and
$b_-$.  The main features of these regimes are:
\begin{itemize} 
\item {\bf Subcritical case.}  Here $b_+< b_-$, or deleterious mutations
  prevail.  The fitness of the population eventually reaches a steady state.
  
\item {\bf Critical case.}  Here $b_+= b_-$, or no mutational bias.  The
  average fitness of the population grows as $t^{2/3}$ and the width of the
  fitness distribution grows as $t^{1/2}$.
  
\item {\bf Supercritical case.}  Here $b_+> b_-$, or advantageous mutations
  are favorable.  The average fitness grows linearly in time and the width of
  the fitness distribution still grows as $t^{1/2}$.
\end{itemize}

In all three cases, the total population density $N$ and the average age $A$,
defined by
\begin{equation}
  \label{adef}
A  = {1\over N}\sum_{n=1}^\infty \int da\,a\,C_n(a)
\end{equation}
saturate to finite values.  The steady state values of $N$ and $A$ may be
determined by balance between the total birth rate $B\equiv b+ b_++ b_-$ and
the death rate $\gamma N$ due to overcrowding.  For example, in the critical
and supercritical cases, there are essentially no individuals with small
fitness, so that the last two terms in Eq.~(\ref{N}) may be neglected.  Then
the steady-state solution to this equation is simply
\begin{equation}
\label{N-ss}
N={B \over \gamma}.
\end{equation}
This statement also expresses the obvious fact that in the steady state the
total birth rate $B$ must balance the total death rate $\gamma N$.  (For fit
populations, the death rate due to aging is negligible.)~ Similarly, the
average age may be inferred from the condition it must equal the average time
between death events.  Thus
\begin{equation}
\label{A-ss}
A={1\over \gamma N}={1\over B}.
\end{equation}
The behavior of the average age in the subcritical case is more subtle and we
treat this case in detail in the section following.

\section{the subcritical case}

When deleterious mutations are favored ($b_->b_+$), the random-walk-like
master equation for $P_n$ contains both the mutational bias towards the
absorbing boundary at the origin, as well as an effective positive bias due
to the $1/n$ term on the right-hand side of Eq.~(\ref{pn}).  The balance
between these two opposite biases leads to a stationary state whose solution
is found by setting $\dot P_n=0$ in Eq.~(\ref{pn}).  To obtain this steady
state solution, it is convenient to introduce the generating function
\begin{equation}
  \label{GF-def}
F(x)=\sum_{n=1}^\infty P_n\,x^{n-1}.  
\end{equation}
Multiplying Eq.~(\ref{pn}) by $x^{n-1}$ and summing over $n$ gives
\begin{equation}
\label{GF-int}
0=(b-\gamma N)F-\sum_{n=1}^\infty {P_n\over n}x^{n-1}+b_+xF+b_-\left({F\over x}-P_1\right).
\end{equation}
The term involving $P_n/n$ is simplified by using
\begin{equation}
\label{GF-inv}
\sum_{n=1}^\infty{P_n\over n}x^{n-1} ={1\over x}\int_0^x F(y)\, dy.
\end{equation}
Multiplying Eq.~(\ref{GF-int}) by $x$ and differentiating with respect to $x$
gives
\begin{equation}
\label{f2}
{F'(x)\over F(x)}={\gamma N-b+1-2 b_+x\over b_+x^2 -(\gamma N-b)x
+ b_-},
\end{equation}
where the prime denotes differentiation.

As in the case of the master equation for $P_n$, this differential equation
for $F$ has a hidden indeterminacy, as the total population density
$N=F(x=1)$ appears on the right-hand side.  Thus an integration of
Eq.~(\ref{f2}), subject to the boundary condition $F(1)=N$, actually gives a
family of solutions which are parameterized by the value of $N$.  While the
family of solutions can be obtained straightforwardly by a direct integration
of Eq.~(\ref{f2}), only one member of this family is the correct one.
To determine this true solution, we must invoke additional arguments about
the physically realizable value of $N$ for a given initial condition.

An upper bound for $N$ may be found from the steady-state version of
Eq.~(\ref{N}),
\begin{equation}
  \label{ME-ss}
(B-\gamma N)N= \sum_{n=1}^\infty {P_n\over n} + b_-P_1.
\end{equation}
Since the right-hand side must be non-negative, this provides the bound
$\gamma N<B$.  On the other hand, we may obtain a lower bound for $N$ by
considering the master equation for $P_n$ in the steady state.  For
$n\to\infty$, we may neglect the $P_n/n$ term in Eq.~(\ref{pn}) and then
solve the resulting equation to give
$P_n=A_+\lambda_+^n+A_-\lambda_-^n$, where
\begin{equation}
  \label{lambda}
\lambda_\pm=\left[\gamma N-b\pm\sqrt{(\gamma N-b)^2-4 b_+b_-}\right]/ 2 b_-.
\end{equation}
For $P_n$ to remain positive, $\lambda_\pm$ should be real. Hence we require
$\gamma N>b+2\sqrt{b_+b_-}$.  We therefore conclude
that $N$ must lie in the range
\begin{equation}
\label{bounds}
b+2\sqrt{b_+b_-}\leq \gamma N<B.
\end{equation}

\begin{figure}
\narrowtext
\epsfxsize=60mm\epsfysize=60mm
\hskip 0.25in\epsfbox{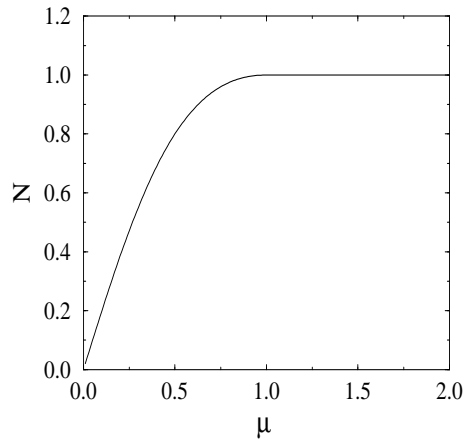}
\vskip 0.1in
\caption{Minimal steady state value of the total density $N$ versus mutational 
  bias $\mu= \sqrt{b_+/b_-}$.  Here $\gamma=1$, $b=0$, and $b_++b_-=1$,
  so that the total birth rate $B=b+b_++b_-$ is fixed.  For $\mu >1$, $N$
  sticks at the value of unity.
\label{Ne}}
\end{figure}

While the foregoing suggests that $N$ lies within a finite range, we find
numerically that the minimal solution, which satisfies the lower bound of
Eq.~(\ref{bounds}), is the one that is generally realized (Fig.~\ref{Ne}).
This selection phenomenon is reminiscent of the corresponding behavior in the
Fisher-Kolmogorov equation and related reaction-diffusion
systems\cite{murray}, where only the extremal solution is selected from a
continuous range of {\it a priori} solutions.

To understand the nature of this extremal solution in the present context,
notice that with the bounds on $N$ given in Eq.~(\ref{bounds}), $\lambda_+$
lies within the range $[\mu,1)$, where
\begin{equation}
\label{mu-def}
\mu\equiv\sqrt{b_+\over b_-}
\end{equation}
is the fundamental parameter which characterizes the mutational bias.
Consequently the steady-state fitness distribution decays exponentially with
$n$, namely $P_n\sim \lambda_+^n$.  When the total population density attains
the minimal value $N_{\rm min}= (b+2\sqrt{b_+ b_-})/\gamma$, $\lambda_+$ also
achieves its minimum possible value $\lambda_+^{\rm min}=\mu$, so that the
fitness distribution has the most rapid decay in $n$ for the minimal
solution.  Based on the analogy with the Fisher-Kolmogorov
equation\cite{murray}, we infer that there are two distinct steady-state
behaviors for $P_n$ as a function of the initial condition $P_n(0)$.  For any
$P_n(0)$ with either a finite support in $n$ or decaying at least as fast as
$\mu^{n}$, the extremal solution $P_n\sim \mu^{n}$ is approached as
$t\to\infty$.  Conversely, for initial conditions in which $P_n(0)$ decays
more slowly than $\mu^n$, for example as $\alpha^n$, with $\alpha$ in the
range $(\mu,1)$, the asymptotic solution also decays as $\alpha^n$.
Correspondingly, Eq.~(\ref{pn}) in the steady state predicts a larger than
minimal population density $N=(b+ b_-\alpha+ b_+\alpha^{-1})/\gamma$.

\begin{figure}
\narrowtext
\epsfxsize=60mm\epsfysize=60mm
\hskip 0.25in\epsfbox{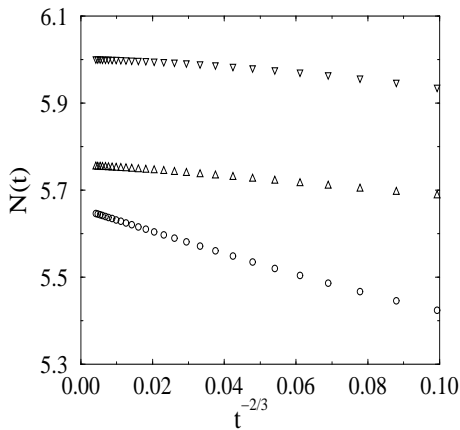}
\vskip 0.1in
\caption{$N(t)$ versus $t^{-2/3}$ in the subcritical case for 
  $b=0$, $b_+=1$, $b_-=2$, and $\gamma=0.5$ for the initial conditions: (i)
  $P_n(t=0)=0.1$ for $1\leq n\leq 10$ ($\circ$), (ii) $P_n(0)=\alpha^n$, with
  $\alpha=(1+\mu)/2$ ($\Delta$), and (iii) $P_n(0)=1/n^2$ for ($\nabla$).
  Asymptotically, the data for $N(t)$ approach the respective theoretical
  values of $N_{\rm min}=4\sqrt{2}\approx 5.6568$,
  $N(\infty)=(2\alpha+\alpha^{-1})/\gamma\approx 5.7574$, and
  $N(\infty)=B/\gamma=6$.  The rate of approach is $t^{-2/3}$ in the first
  case and faster than $t^{-2/3}$ in the latter two cases.  These
  calculations use the full machine precision of $10^{-308}$.
\label{Nt}}
\end{figure}

We also find that the extremal and the non-extremal solutions exhibit
different relaxations to the steady state.  For those initial conditions
which evolve to the extremal solution, the deviation of $N$ and indeed each
of the $P_n$ from their steady state values decay as $t^{-2/3}$, while for
all other initial conditions, the relaxation to the steady state appears to
follow a $t^{-1}$ power law decay.  The power-law approach to the steady
state is surprising, since the overall density obeys a logistic-like
dynamics, $\dot N=bN-\gamma N^2$, for which the approach to the steady state
is exponential.  These results are illustrated in Fig.~\ref{Nt} which shows
the asymptotic time dependence of $N(t)$ based on a numerical integration of
Eq.~(\ref{pn}) with the fourth-order Runge-Kutta algorithm\cite{nr}.  The
demonstration of the $t^{-2/3}$ relaxation to the extremal solution relies on
a correspondence to the transient behavior in the critical case.  This is
presented in Appendix B.

\begin{figure}
\narrowtext
\epsfxsize=60mm\epsfysize=60mm
\hskip 0.25in\epsfbox{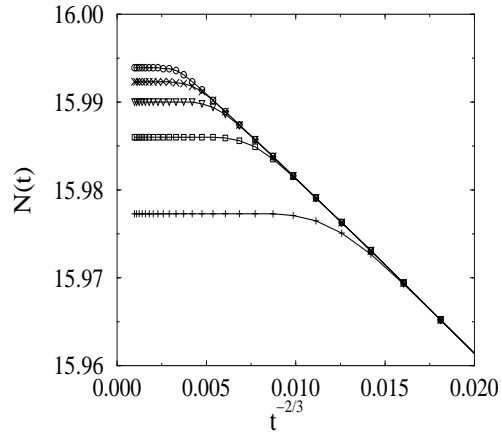}
\vskip 0.1in
\caption{Behavior of $N(t)$ versus $t^{-2/3}$ for precision equal to
  $10^{-100}$, $10^{-150}$, $10^{-200}$, $10^{-250}$, and $10^{-308}$ (bottom
  to top) for the case $b_+=16$, $b_-=1$, $b=0$, and $\gamma=0.5$.  In the
  limit of infinite precision $N(t)\to 16$ as $t\to \infty$.
\label{cutoff}}
\end{figure}

A disconcerting feature of the numerical calculation for $N(t)$ is the small
disagreement between the numerically observed values of the steady-state
population density and the expected theoretical prediction
(Fig.~\ref{cutoff}).  This discrepancy arises from the finite computer
precision which causes very small values of $P_n$ to be set to zero.  To
confirm this, we changed the computer precision from $10^{-100}$ to the full
machine precision of $10^{-308}$ (Fig.~\ref{cutoff}).  As the precision is
increased, $N$ saturates to progressively higher values and approaches the
theoretical prediction.  A similar precisions-dependent phenomenon has been
observed in the context of traveling Fisher-Kolmogorov wave
propagation\cite{derrida,kns}.
 
For the relevant situation where the density $N$ takes the minimal value, we
may rewrite Eq.~(\ref{f2}) as
\begin{equation}
\label{GF-min}
{F'\over F} ={2\mu \over 1-\mu x}+{1\over b_- (1-\mu x)^2}
\end{equation}
Integrating from $x=1$ to $x$ and using $F(1)=N$ gives
\begin{equation}
\label{f3}
F(x)=N\left({1-\mu \over 1-\mu x}\right)^2 
\exp\left\{{1\over b_-\mu}\left({1\over 1-\mu x}-{1\over 1-\mu}\right)
\right\}.
\end{equation}
One can now formally determine $P_n$ by expanding $F(x)$ in a Taylor
series.  For example,
\begin{eqnarray*}
P_1&=&N (1-\mu)^2\,\exp\left\{-{1\over b_-(1-\mu)}\right\},\\
P_2&=&N (1-\mu)^2 \left(2\mu+{1\over b_-}\right)\,
\exp\left\{-{1\over b_-(1-\mu)}\right\}.
\end{eqnarray*}
For many applications, however, there is no need to deal with these unwieldy
expressions.  As we now discuss, the overall fitness or age characteristics
of the population can be obtained directly from the generating function
without using the explicit formulae for the $P_n$.

\subsection{Fitness characteristics} 

Consider the average fitness of the population
\begin{equation}
\label{avfdef}
\langle n\rangle ={1\over N}\sum_{n=1}^\infty n\,P_n,
\end{equation}
which can be expressed in terms of the generating function as
\begin{equation}
\label{avf-GF}
\langle n\rangle ={1\over N}\,{dF\over dx}\bigg|_{x=1}+1.
\end{equation}
From Eq.~(\ref{f3}) we thereby obtain the average fitness
\begin{equation}
\label{avf}
\langle n\rangle ={2\mu \over 1-\mu}+{1\over  b_-(1-\mu)^2}+1.
\end{equation}
As one might anticipate, the average fitness diverges as $\mu\to 1$ from
below, corresponding to the population becoming mutationally neutral.  To
determine the dispersion of the fitness distribution we make use of the
relation
\begin{equation}
\label{nn1}
\langle n(n-1)\rangle ={1\over N}\sum_{n=1}^\infty n(n-1)\,P_n
={1\over N}\,{d^2{(xF)}\over dx^2}\bigg|_{x=1}.
\end{equation}
Substituting Eqs.~(\ref{f3}) and also Eq.~(\ref{avf}) then gives
\begin{eqnarray*}
\langle n^2\rangle =1+{6\mu \over (1-\mu)^2}+{3(1+\mu)\over  b_-(1-\mu)^3}
+{1\over  b_-^2(1-\mu)^4}.
\end{eqnarray*}
Thus the dispersion $\sigma^2=\langle n^2\rangle -\langle n\rangle^2$ in the
fitness distribution is
\begin{equation}
\label{sigma}
\sigma^2={2\mu\over (1-\mu)^2}
+{\mu+1\over  b_-(1-\mu)^3}.
\end{equation}
As the mutational bias vanishes, $\mu\to 1$, the average fitness and the
dispersion diverge as $\langle n\rangle\simeq b_-^{-1}(1-\mu)^{-2}$ and
$\sigma\simeq \sqrt{2/ b_-}(1-\mu)^{-3/2}$.  Thus these two moments are
related by $\sigma\sim \langle n\rangle^{3/4}$.  As we shall see in Sec. VI,
this basic relation continues to hold in the critical case.

\subsection{Age characteristics} 

In the steady state, we solve Eq.~(\ref{cn}) to give the concentration of
individuals with age $a$ and fitness $n$
\begin{equation}
\label{cnasol}
C_n(a)=P_n\left(\gamma N+{1\over n}\right)\,
\exp\left[-\left(\gamma N+{1\over n}\right)a\right].
\end{equation}
The average age of the population is
\begin{eqnarray}
\label{A-def}
A&=&{1\over N}\sum_{n=1}^\infty \int_0^\infty da\,a\,C_n(a)\nonumber \\
 &=&{1\over N}\sum_{n=1}^\infty {P_n\over \gamma N+n^{-1}},
\end{eqnarray}
where the second line is obtained by using Eq.~(\ref{cnasol}).  This
expression can be rewritten directly in terms of the generating function by
first noticing that
\begin{eqnarray}
\label{GF-rel}
\int_0^1 x^\nu F(x)\, dx &=& \int_0^1 \sum_{n=1}^\infty P_n x^{n+\nu-1}\,dx\nonumber\\
    &=& \sum_{n=1}^\infty {P_n\over n+\nu}.
\end{eqnarray}
Thus we re-express Eq.~(\ref{A-def}) in a form which allows us to exploit
Eq.~(\ref{GF-rel}).  After several elementary steps, we obtain
\begin{eqnarray}
\label{age}
A &=& {1\over \gamma N}-{1\over N}\,{1\over (\gamma N)^2}
      \sum_{n=1}^\infty{P_n\over n+(\gamma N)^{-1}}\nonumber \\
  &=& {1\over \gamma N}-{1\over N}\,{1\over (\gamma N)^2}
     \int_0^1 dx\,x^{1\over \gamma N}\,F(x).
\end{eqnarray}

\begin{figure}
\narrowtext
\epsfxsize=60mm\epsfysize=50mm
\hskip 0.4in\epsfbox{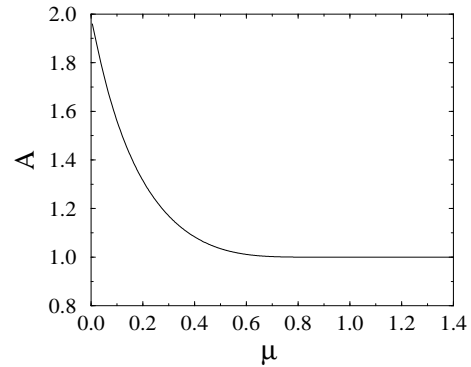}
\vskip 0.1in
\caption{Average age $A$ of the steady state population versus mutational
  bias $\mu=\sqrt{b_+/b_-}$.  The total birth rate $B=b+b_++b_-$ is held
  fixed throughout, with $b=0$, $b_++b_-=1$, and $\gamma$ given by the
  extremal steady-state solution $\gamma N=b+2\sqrt{b_+b_-}$.  For $\mu>1$,
  the average age sticks at the value of unity, while for $\mu\to 0$, $A\to
  2$.
\label{ava}}
\end{figure}

This expression should be compared with the result for the critical and
supercritical cases, namely $A=(\gamma N)^{-1}=B^{-1}$ (see
Eq.~(\ref{A-ss})).  In the subcritical case, $\gamma N <B$ and the above two
expressions $A_{\rm min}=B^{-1}$ and $A_{\rm max}=(\gamma N)^{-1}$ provide
lower and upper bounds for the average age.  This is proved in Appendix A.
Fig.~\ref{ava} shows the surprising feature of Eq.~(\ref{age}) that the
average age {\em decreases} as the population gets fitter!  We also see that
the average age of the least fit population $(\mu\to 0)$ is twice that of the
increasingly fit populations in the critical and supercritical cases.  We now
demonstrate this fact.  To provide a fair comparison we take the total birth
rate rate to be equal to unity in both cases and also choose $b=0$ for
simplicity.  For fit populations (critical and supercritical cases), the
average age is simply $A=B^{-1}=1$.  For the least fit population $\mu\to 0$,
and correspondingly $N\to 0$.  In this limit, we may write Eq.~(\ref{A-def})
as,
\begin{equation}
\label{A-der}
A = {1\over N}\sum_{n=1}^\infty{P_n\over \gamma N+n^{-1}} 
  \approx {1\over N} \sum_{n=1}^\infty nP_n 
    =\langle n\rangle.
\end{equation}
On the other hand, from Eq.~(\ref{avf}) we have
\begin{equation}
\langle n\rangle \approx 1+{1\over b_-}\approx 2.
\end{equation}

The relation $A=\langle n\rangle$ is natural for the least fit population, as
the total density is small and competition among individuals plays an
insignificant role.  Thus the average age may be found by merely averaging
the intrinsic life expectancy of the population.  Intriguingly, in this limit
the average individual in the least fit population lives twice as long as
individuals in relatively fit populations.  

It is also worth noting that in the limit of a minimally fit population
($\mu\to 0$) we can expand the generating function in Eq.~(\ref{f3})
systematically.  We thereby find that the density $P_n$ exhibits a
super-exponential decay, $P_n=Ne^{-1}/(n-1)!$.

\section{the supercritical case}

When advantageous mutations are favored, the master equation for $P_n$,
Eq.~(\ref{pn}), can be viewed as a random walk with a bias towards
increasing $n$.  Because there is no mechanism to counterbalance this bias,
the average fitness grows without bound and no steady state exists.  As in
the case of a uniformly biased random walk on a semi-infinite domain, the
distribution of fitness becomes relatively localized in fitness space, with
the peak drifting towards increasing $n$ with a velocity $V= b_+- b_-$.
Since the fitness profile is non-zero only for large $n$ in the long time
limit, it is appropriate to adopt continuum approach.  We therefore treat $n$
as continuous, and derive the continuum limits of Eqs.~(\ref{pn}) and
(\ref{N}).  For the time evolution of the fitness distribution $P(n,t)$, we
obtain the equation of motion
\begin{equation}
\label{px}
\left({\partial \over \partial t}+V{\partial \over \partial n}\right)P
=\left(B-\gamma N-{1\over n}\right)P 
+D{\partial^2 P\over \partial n^2}.
\end{equation}
This is just a convection-diffusion equation, supplemented by a birth/death
term.  Here the difference between advantageous and deleterious mutations
provides the drift velocity $V= b_+- b_-$, and the average mutation rate $D=
(b_++ b_-)/2$ plays the role of diffusion constant.  For the total population
density we obtain
\begin{equation}
\label{Nx}
{d N\over dt}=(B-\gamma N)N-\int_{0}^\infty dn\,{P(n,t)\over n}.
\end{equation}

To determine the asymptotic behavior of these equations, we use the fact that
the fitness distribution becomes strongly localized about a value of $n$
which increases as $Vt$.  Thus we replace the integral in Eq.~(\ref{Nx}) by
its value at the peak of the distribution, $N/Vt$.  With this crude
approximation, Eq.~(\ref{Nx}) becomes
\begin{equation}
\label{Nappr}
{d N\over dt}=\left(B-\gamma N-{1\over Vt}\right)N.
\end{equation}
Thus we conclude that the density approaches its steady state value as
\begin{equation}
\label{Nasymp}
\gamma N\to B-{1\over Vt}.
\end{equation}
This provides both a proof Eq.~(\ref{N-ss}), as well as the rate of
convergence to the steady state.

We now substitute this asymptotic behavior for the total population density
into Eq.~(\ref{px}) to obtain
\begin{equation}
\label{px1}
\left({\partial \over \partial t}+V{\partial \over \partial n}\right)P
=\left({1\over Vt}-{1\over n}\right)P 
+D{\partial^2 P\over \partial n^2}.
\end{equation}
Notice that the birth/death term on the right hand side is negative
(positive) for subpopulations which are less (more) fit than average fitness
$Vt$.  This birth/death term must also be zero, on average, since the total
population density saturates to a constant value.  Moreover, this term must
be small near the peak of the fitness distributions where $n\sim Vt$.  Thus
as a simple approximation, we merely neglect this birth/death term and check
the validity of this assumption {\it a posteriori}.  This transforms
Eq.~(\ref{px1}) into the classical convection-diffusion equation whose
solution is
\begin{equation}
\label{Pappr}
P(n,t)={N\over \sqrt{4\pi D t}}\, \exp\left[-{(n-Vt)^2\over 4 D t}\right].
\end{equation}
This basic results implies that the fitness distribution indeed is a
localized peak, with average fitness growing linearly in time, $\langle
n\rangle=Vt$, and width growing diffusively, $\sigma=\sqrt{2Dt}$.  We now
check the validity of dropping the birth/death term in Eq.~(\ref{px1}).  Near
the peak, $|n-Vt|\sim \sqrt{D t}$, so that the birth/death term is of order
$t^{-3/2}\times P$.  On the other hand, the other terms in Eq.~(\ref{px1})
are of order $t^{-1}\times P$, thus justifying the neglect of birth/death
term.

We now turn to the age characteristics.  Asymptotically, the density of
individuals with given age and fitness changes slowly with time because the
overall density reaches a steady state.  Consequently, the time variable $t$
is {\it slow} while the age variable $a$ is {\em fast}.  Physically this
contrast reflects the fact that during the lifetime of an individual the
change in the overall age characteristics of the population is small.  Thus
in the first approximation, we retain only the age derivative in
Eq.~(\ref{cn}).  We also ignore the term $C_n/n$, which is small near the
peak of the asymptotic fitness distribution.  Solving the resulting master
equation and using the boundary condition of Eq.~(\ref{boundary}) we obtain
\begin{eqnarray}
\label{cnat}
C_n(a,t)&\simeq&P_n(t)\gamma N\,e^{-\gamma Na}\nonumber \\
&=&{\gamma N^2\over \sqrt{4\pi D t}}\,
\exp\left[-\gamma Na-{(n-Vt)^2\over 4D t}\right].
\end{eqnarray}
Integrating over the fitness variable, we find that the age distribution
$C(a,t)=\int dn\, C_n(a,t)$ has the stationary Poisson form
\begin{equation}
\label{cat}
C(a)=\gamma N^2\,e^{-\gamma Na}.
\end{equation}
From this, the average age is $A=(\gamma N)^{-1}=B^{-1}$ in agreement with
Eq.~(\ref{A-ss}).  As discussed in Sec.~III~B, the surprising conclusion is
that the average age in the supercritical case is always {\it smaller} than
that in the subcritical case.

\section{the critical case}

We now consider the critical case where the rates of advantageous and
deleterious mutations are equal.  Without the $1/n$ term and with $\gamma N
-b = b_++b_-$, Eq.~(\ref{pn}) becomes the master equation for an unbiased
random walk on the semi-infinite range $n\ge 0$.  Due to the $1/n$ term, the
system has a bias towards increasing $n$ which vanishes as $n\to\infty$ (see
Fig.~\ref{Pn}).  Thus we anticipate that the average fitness will grow faster
than $t^{1/2}$ and slower than $t$.  Hence we can again employ the continuum
approach to account for the evolution of the $P_n$.  In this limit, the
corresponding master equation for $P(n,t)$ becomes
\begin{equation}
\label{p}
{\partial P\over \partial t}
=\left(B-\gamma N-{1\over n}\right)P 
+D{\partial^2 P\over \partial n^2}.
\end{equation}
Numerically, we find $\langle n\rangle\sim t^{2/3}$, while the dispersion
still grows as $t^{1/2}$, that is, as $\sigma\sim \sqrt{t}$.  Thus these two
quantities are related by $\sigma\sim \langle n\rangle^{3/4}$, as derived
analytically for the subcritical case.

To provide a more quantitative derivation of the above scaling laws for
$\langle n\rangle$ and $\sigma$, as well as to determine the fitness
distribution itself, we examine the equation for $P(n,t)$.  First note that
the total population density still obeys Eq.~(\ref{Nx}), as in the
supercritical case.  Under the assumption that the fitness distribution is
relatively narrow compared to its mean position, a result which we have
verified numerically, we again estimate the integral on the right-hand side
of Eq.~(\ref{Nx}) to be of the order of $N/\langle n\rangle$.  This leads to
\begin{equation}
\label{Nas}
\gamma N\to B-{1\over \langle n\rangle}.
\end{equation}
Substituting this into Eq.~(\ref{p}) yields
\begin{equation}
\label{p1}
{\partial P\over \partial t}=\left({1\over \langle n\rangle}-
{1\over n}\right)P +D{\partial^2 P\over \partial n^2}.
\end{equation}
Given that the peak of the distribution is located near $n\approx \langle
n\rangle$, it proves useful to change variables from $(n,t)$ to the comoving
co-ordinates $(y=n-\langle n\rangle,t)$ to determine how the peak of the
distribution spreads.  We therefore write the derivatives in the comoving
coordinates
\begin{eqnarray*}
{\partial \over \partial t}=
{\partial \over \partial t}\bigg|_{\rm comov.}\!\!\!\!\!\!
-{d \langle n\rangle\over dt}\,{\partial \over \partial y},\qquad
{\partial \over \partial n}={\partial \over \partial y},
\end{eqnarray*}
and expand the birth/death term in powers of the deviation $y=n-\langle
n\rangle$
\begin{eqnarray*}
{1\over \langle n\rangle}-{1\over n}=
{y\over \langle n\rangle^2}-{y^2\over \langle n\rangle^3}+\ldots
\end{eqnarray*}
Now Eq.~(\ref{p1}) becomes
\begin{equation}
\label{p2}
{\partial P\over \partial t}-{d \langle n\rangle\over dt}\,
{\partial P\over \partial y}
={y\over \langle n\rangle^2}\,P-{y^2\over \langle n\rangle^3}\,P 
+D{\partial^2 P\over \partial y^2}.
\end{equation}

Let us first assume that the average fitness grows faster than diffusively,
that is, $\langle n\rangle\gg \sqrt{t}$.  With this assumption, the dominant
terms in Eq.~(\ref{p2}) are
\begin{equation}
\label{p3}
{d \langle n\rangle\over dt}\,{\partial P\over \partial y}
=-{y\over \langle n\rangle^2}\,P.
\end{equation}
These terms balance when $\langle n\rangle(ty)^{-1}\sim y\langle
n\rangle^{-2}$.  Using this scaling and balancing the remaining sub-dominant
terms in Eq.~(\ref{p2}) gives $y\sim \sqrt{t}$.  The combination of these
results yields $\langle n\rangle\sim t^{2/3}$. This justifies our initial
assumption that $\langle n\rangle\gg \sqrt{t}$.  Now we write $\langle
n\rangle=(ut)^{2/3}$, with $u$ of order unity, to simplify Eq.~(\ref{p3}) to
\begin{equation}
\label{p4}
{\partial P\over \partial y}
=-{3y\over 2u^2 t}\,P.
\end{equation}
In terms of $n=y+\langle n\rangle$ the solution is
\begin{equation}
\label{p5}
P(n,t)=N\sqrt{3\over 4\pi u^2 t}\,
\exp\left\{-{3\left[n-(ut)^{2/3}\right]^2\over 4u^2 t}\right\}.
\end{equation}
Thus the fitness distribution is again Gaussian, as in the supercritical
case, but with the average fitness growing as $\langle n\rangle=(ut)^{2/3}$.
Finally, we determine $u=\sqrt{3D}$ by substituting $\langle
n\rangle=(ut)^{2/3}$ in Eq.~(\ref{p2}) and balancing the sub-dominant terms.

The age distribution in the critical case can be obtained in similar manner
as in the supercritical case.  The approximations that were invoked to
determine the age distribution in the supercritical case still apply.
Consequently, the asymptotic form for $C_n(a)$ is still given by
Eq.~(\ref{cnat}), and this gives the same expression for $C(a)$ after
integrating over $n$, as in Eq.~(\ref{cat}).  Hence the average age is again
$B^{-1}$, as in Eq.~(\ref{A-ss}).


\section{summary and discussion}

We have introduced an age-structured logistic-like population dynamics model,
which is augmented by fitness mutation of offspring with respect to their
parents.  Here fitness is quantified by the life expectancy $n$ of an
individual.  We found unusual age and fitness evolution in which the overall
mutational bias leads to three distinct regimes of behavior.  Specifically,
when deleterious mutations are more likely, the fitness distribution of the
population approaches a steady state which is an exponentially decaying
function of fitness.  When advantageous mutations are favored or when there
is no mutational bias, a Gaussian fitness distribution arises, in which the
average fitness grows as $\langle n\rangle= Vt$ and as $\langle
n\rangle=(3D)^{1/3}t^{2/3}$, respectively.

Paradoxically, the average age of the population is maximal for a completely
unfit population.  Conversely, individuals are less long-lived for either
positive or no mutational bias, even though the average fitness increases
indefinitely with time.  That is, a continuous ``rat-race'' towards increased
fitness leads to a {\em decrease} in the average life span.  As individuals
become fit, increased competition results in their demise well before their
intrinsic lifetimes are reached.  Thus within our model, an increase in the
average fitness is not a feature which promotes longevity.

Our basic conclusions should continue to hold for the more realistic
situation where the mortality rate increases with age
\cite{charlesworth,evolution,azbel}.  The crucial point is that old age is
unattainable within our model, even if individuals are infinitely fit.  When
the mutational bias is non-negative, old age is unattainable due to keen
competition among fit individuals, while if deleterious mutations are
favored, age is limited by death due to natural mortality.  In either case,
there are stringent limits on the life expectancy of any individual.  To
include an age-dependent mortality into our model, we may write the mortality
term $f_n(a)C_n(a,t)$ instead of $n^{-1}C_n(a,t)$ in Eq.~(\ref{cn}), where
$f_n(a)$ is the mortality rate for individuals of age $a$.  Similarly, the
term $n^{-1}P_n$ in Eq.~(\ref{pn}) should be replaced by $\int da
f_n(a)C_n(a,t)$.  However, these generalized terms play no role for large
$n$, since $f_n(a)$ is a decreasing function of $n$ and old age is
unattainable.

\smallskip
We gratefully acknowledge partial support from NSF grant DMR9632059 and ARO
grant DAAH04-96-1-0114.

\end{multicols}

\widetext
\appendix

\section{Bounds for the average age}
 
The upper bound, $A<(\gamma N)^{-1}$, follows immediately from
Eq.~(\ref{age}), so we just prove $A> A_{\rm min}$.  We have
\begin{eqnarray*}
A_{\rm min}&=&B^{-1}={1\over b+ b_-(1+\mu^2)}\\
A&=&{1\over b+2 b_-\mu}-{1\over(b+2 b_-\mu)^2}\,
\int_0^1 dx\,x^{1\over b+2 b_-\mu}
\left({1-\mu\over 1-\mu x}\right)^2 \exp\left\{{1\over  b_-\mu}
\left({1\over 1-\mu x}-{1\over 1-\mu}\right)\right\}.
\end{eqnarray*}
Using these expressions and performing elementary transformations we reduce
the inequality $A>A_{\rm min}$ to
\begin{equation}
\label{1}
\int_0^1 {dx\over  b_-(1-\mu x)^2}\,
x^{1\over b+2 b_-\mu}\exp\left\{{1\over  b_-\mu}
\left({1\over 1-\mu x}-{1\over 1-\mu}\right)\right\}
<{b+2 b_-\mu\over b+ b_-(1+\mu^2)}.
\end{equation}
Let us now introduce the variable
\begin{equation}
\label{2}
v=-{1\over b_-\mu}
\left({1\over 1-\mu x}-{1\over 1-\mu}\right),
\end{equation}
so that $dv=dx/b_-(1-\mu x)^2$, which varies in the range $[0,V]$, with
$V={1\over b_-(1-\mu)}$.  This simplifies Eq.~(\ref{1}) to
\begin{equation}
\label{3}
\int_0^V dv\,e^{-v}
\left[{1-V^{-1}v\over 1-V^{-1}\mu v}\right]
^{1\over b+2 b_-\mu}
<{b+2 b_-\mu\over b+ b_-(1+\mu^2)}.
\end{equation}
We now use the inequality
\begin{equation}
\label{4}
\left[{1-p\over 1-q}\right]^\nu<e^{(q-p)\nu}
\end{equation}
which holds for $0<q<p<1$ and $\nu>0$.  This inequality is readily proven by
taking the logarithm on both sides and using the expansion $\ln(1-u)=
-\sum_{k\geq 1} u^k/k$.  Now we apply Eq.~(\ref{4}) to the integrand in
(\ref{3}) and then replace the upper limit $V$ in the integral by
$\infty$ to give
\begin{eqnarray*}
\int_0^V dv\,e^{-v}\left[{1-V^{-1}v\over 1-V^{-1}\mu v}\right]
^{1\over b+2 b_-\mu}&<&\int_0^V dv\,
\exp\left\{-v-v{ b_-(1-\mu)^2\over b+2 b_-\mu}\right\}\\
&<&\int_0^\infty dv\,
\exp\left\{-v{b+ b_-(1+\mu^2)\over b+2 b_-\mu}\right\}\\
&=&{b+2 b_-\mu\over b+ b_-(1+\mu^2)}.
\end{eqnarray*}
This completes the proof. 

The lower bound $A_{\rm min}$ turns out to be very accurate in the case when
mutations are slightly deleterious.  To see this let us write $b_+=1,
b_-=(1+\epsilon)^2$, where $\epsilon\ll 1$.  Replacing $x$ by the transformed
variable $v=\epsilon^{-1}-(1+\epsilon-x)^{-1}$ recasts the integral
Eq.~(\ref{age}) as
\begin{eqnarray}
\label{J}
\epsilon^2\int_0^{1\over\epsilon(1+\epsilon)} dv\,e^{-v}
\left(1-{\epsilon^2 v\over 1-\epsilon v}\right)^{1\over b+2+2\epsilon}.
\end{eqnarray}
We now expand the integrand, 
\begin{eqnarray*}
\left(1-{\epsilon^2 v\over 1-\epsilon v}\right)^{1\over b+2+2\epsilon}=
1-{\epsilon^2 v\over b+2+2\epsilon}
-{\epsilon^3 v^2\over b+2+2\epsilon}+{\cal O}(\epsilon^4),
\end{eqnarray*}
replace the upper limit in the integral Eq.~(\ref{J}) by $\infty$, and
compute the resulting simple integrals explicitly to obtain a series
expansion in $\epsilon$ for the average age.  Together with analogous
expansions for $A_{\rm max}$ and $A$ we have
\begin{eqnarray}
A_{\rm max}&=&{1\over b+2+2\epsilon}\nonumber \\
A_{\rm min}&=&A_{\rm max}-{\epsilon^2\over (b+2+2\epsilon)^2}
+{\epsilon^4\over (b+2+2\epsilon)^3}+{\cal O}(\epsilon^6)\nonumber\\
A&=&A_{\rm max}-{\epsilon^2\over (b+2+2\epsilon)^2}
+{\epsilon^4\over (b+2+2\epsilon)^3}
+{2\epsilon^5\over (b+2+2\epsilon)^3}+{\cal O}(\epsilon^6).
\end{eqnarray}
Thus the difference between the exact value and $A_{\rm min}$ is of order
$\epsilon^5$.

\section{Transient behavior of the total density}

Numerically, we found that in the subcritical case the total population
density approaches the steady state value $N_\infty=
(b+2\sqrt{b_+b_-})/\gamma$ from below with a deviation that vanishes as
$t^{-2/3}$.  We now explain this behavior by constructing a mapping between
this transient behavior in the subcritical case and the transient behavior in
the critical case.  We start with the basic rate equation, Eq.~(\ref{pn}).
We may remove the term $\gamma NP_n$ through the transformation
\begin{equation}
\label{qpn}
Q_n(t)=P_n(t)\exp\left\{\gamma\int_0^t dt'\,N(t')\right\},
\end{equation}
which simplifies Eq.~(\ref{pn}) to
\begin{equation}
\label{qn}
{d Q_n\over dt}=\left(b-{1\over n}\right)Q_n 
+ b_+Q_{n-1}+ b_-Q_{n+1}.
\end{equation}
Next, the steady state behavior $P_n\sim \mu^n$ suggests replacing the
$Q_n$ by $R_n(t)=\mu^{-n} Q_n(t)$.  This also removes the asymmetry in the birth
terms and gives
\begin{equation}
\label{rn}
{d R_n\over dt}=\left(b-{1\over n}\right)R_n 
+b_*(R_{n-1}+R_{n+1}),
\end{equation}
where we use the shorthand notation $b_*=\sqrt{b_+ b_-}$. 

One cannot use the continuum approximation to determine the steady-state
solutions for $P_n$ or $Q_n$.  However, the continuum approximation is
appropriate for the $R_n$.  Then Eq.~(\ref{rn}) reduces to
\begin{equation}
\label{r}
{\partial R\over \partial t}
=\left(b+2b_*-{1\over n}\right)R 
+b_*{\partial^2 R\over \partial n^2},
\end{equation}
which is very similar to Eq.~(\ref{p1}).  Hence we expect that the
distribution of $R_n$ is peaked around $\langle n\rangle\simeq
(3b_*)^{1/3}t^{2/3}$.  It proves convenient to make this scaling manifest.
To this end we change variables once more,
\begin{equation}
\label{srn}
S_n(t)=R_n(t)\exp\left\{-(b+2b_*)t
+\left({9t\over b_*}\right)^{1/3}\right\},
\end{equation}
to get
\begin{equation}
\label{s}
{\partial S\over \partial t}
=\left({1\over \langle n\rangle}-{1\over n}\right)S 
+b_*{\partial^2 S\over \partial n^2}.
\end{equation}
Repeating the procedure of Sec.~V we determine the asymptotic solution to
Eq.~(\ref{s}) as
\begin{equation}
\label{s1}
S_n(t)\sim {1\over \sqrt{4\pi b_* t}}\,
\exp\left\{-{(n-\langle n\rangle)^2\over 4b_* t}\right\}.
\end{equation}

To find the asymptotics of the total population density let us compute
$\sum Q_n(t)$.  First, (\ref{qpn}) can be expressed as
\begin{equation}
\label{qnn}
\sum_{n=1}^\infty Q_n(t)=N(t)\exp\left\{\gamma\int_0^t dt'\,N(t')\right\}.
\end{equation}
On the other hand, 
\begin{equation}
\label{qrs}
\sum_{n=1}^\infty Q_n(t)=\sum_{n=1}^\infty \mu^n R_n(t)
=\exp\left\{(b+2b_*)t-\left({9t\over b_*}\right)^{1/3}\right\}
\sum_{n=1}^\infty \mu^n S_n(t).
\end{equation}
In the last sum, the factor $\mu^n$ suggests that only terms with small $n$
contribute significantly.  Although the asymptotic expression (\ref{s1}) is
formally justified only in the scaling region, where $|n-\langle
n\rangle|\sim \sqrt{b_*t}$, the continuum approach typically provides a
qualitatively correct description even outside this region.  Therefore we
take Eq.~(\ref{s1}) to estimate $S_n$ for small $n$.  We find
\begin{equation}
\label{s2}
\sum_{n=1}^{\infty}\mu^n S_n(t)\sim 
\exp\left\{-C\left({9t\over b_*}\right)^{1/3}\right\},
\end{equation}
where we use $\langle n\rangle \sim t^{2/3}$, as in the critical case, and
$C$ is a constant.  By substituting Eq.~(\ref{s2}) into Eq.~(\ref{qrs}) we
obtain
\begin{equation}
\label{qc}
\sum_{n=1}^\infty Q_n(t)\sim 
\exp\left\{(b+2b_*)t-(1+C)\left({9t\over b_*}\right)^{1/3}\right\}.
\end{equation}
Combining Eqs.~(\ref{qnn}) and (\ref{qc}) we arrive at the asymptotic expansion
\begin{equation}
\label{intN}
\gamma\int_0^t dt'\,N(t')=(b+2b_*)t
-(1+C)\left({9t\over b_*}\right)^{1/3}+\ldots,
\end{equation}
which implies
\begin{equation}
\label{N23}
\gamma N(t)= b+2b_*-{\rm const}\times t^{-2/3}+\ldots
\end{equation}

\end{document}